\documentclass[10pt,letterpaper]{article}
\usepackage[top=0.85in,left=2.75in,footskip=0.75in]{geometry}

\usepackage{amsmath,amssymb}

\usepackage{changepage}

\usepackage{textcomp,marvosym}

\usepackage{cite}

\usepackage{nameref,hyperref}

\usepackage[right]{lineno}

\usepackage[nopatch=eqnum]{microtype}
\DisableLigatures[f]{encoding = *, family = * }

\usepackage[table]{xcolor}

\usepackage{array}

\usepackage{gensymb}

\newcolumntype{+}{!{\vrule width 2pt}}

\newlength\savedwidth

\newcommand\thickhline{\noalign{\global\savedwidth\arrayrulewidth\global\arrayrulewidth 2pt}%
\hline
\noalign{\global\arrayrulewidth\savedwidth}}

\raggedright
\usepackage[aboveskip=1pt,labelfont=bf,labelsep=period,justification=raggedright,singlelinecheck=off]{caption}

\makeatletter
\renewcommand{\@biblabel}[1]{\quad#1.}
\makeatother

\usepackage{lastpage,fancyhdr,graphicx}
\usepackage{epstopdf}
\fancyheadoffset[L]{2.25in}
\fancyfootoffset[L]{2.25in}
\newcommand{\mytexttilde}{\raisebox{0.5ex}{\texttildelow}}

\begin{document}
\vspace*{0.2in}

\begin{flushleft}
{\Large
\textbf\newline{Cardiac digital twins at scale from MRI: Open tools and representative models from \mytexttilde{55000} UK Biobank participants} % Please use "sentence case" for title and headings (capitalize only the first word in a title (or heading), the first word in a subtitle (or subheading), and any proper nouns).
}
\newline
\\
Devran Ugurlu\textsuperscript{1,2,3\Yinyang},
Shuang Qian\textsuperscript{1\Yinyang},
Elliot Fairweather\textsuperscript{1\Yinyang},
Charlene Mauger\textsuperscript{1},
Bram Ruijsink\textsuperscript{1,4},
Laura Dal Toso\textsuperscript{5},
Yu Deng\textsuperscript{1},
Marina Strocchi\textsuperscript{1,2,3},
Reza Razavi\textsuperscript{1},
Alistair Young\textsuperscript{1\ddag},
Pablo Lamata\textsuperscript{1\ddag},
Steven Niederer\textsuperscript{1,2,3\ddag},
Martin Bishop\textsuperscript{1\ddag}
\\
\bigskip
\textbf{1} School of Biomedical Engineering \& Imaging Sciences, King's College London, London, UK
\\
\textbf{2} The Cardiac Electro-Mechanics Research Group, Imperial College London, London, UK
\\
\textbf{3} The Alan Turing Institute, London, UK
\\
\textbf{4} Department of Cardiology, University Medical Center Utrecht, Utrecht, NLD
\\
\textbf{5} ETH Zurich, Zurich, Switzerland
\\
\bigskip

% Insert additional author notes using the symbols described below. Insert symbol callouts after author names as necessary.
% 
% Remove or comment out the author notes below if they aren't used.
%
% Primary Equal Contribution Note
\Yinyang These authors contributed equally to this work.

% Additional Equal Contribution Note
% Also use this double-dagger symbol for special authorship notes, such as senior authorship.
\ddag These authors also contributed equally to this work.

% % Current address notes
% \textcurrency Current Address: Dept/Program/Center, Institution Name, City, State, Country % change symbol to "\textcurrency a" if more than one current address note
% % \textcurrency b Insert second current address 
% % \textcurrency c Insert third current address

% % Deceased author note
% \dag Deceased

% % Group/Consortium Author Note
% \textpilcrow Membership list can be found in the Acknowledgments section.

% Use the asterisk to denote corresponding authorship and provide email address in note below.
* devran.ugurlu@kcl.ac.uk

\end{flushleft}
\section*{Abstract}
A cardiac digital twin is a virtual replica of a patient's heart for screening, diagnosis, prognosis, risk assessment, and treatment planning of cardiovascular diseases. This requires an anatomically accurate patient-specific 3D structural representation of the heart, suitable for electro-mechanical simulations or study of disease mechanisms. However, generation of cardiac digital twins at scale is demanding and there are no public repositories of models across demographic groups. We describe an automatic open-source pipeline for creating patient-specific left and right ventricular meshes from cardiovascular magnetic resonance images, its application to a large cohort of \mytexttilde{55k} participants from UK Biobank, and the construction of the most comprehensive cohort of adult heart models to date, comprising 1423 representative meshes across sex (male, female), body mass index (range: 16 - 42 kg/m\textsuperscript{2}) and age (range: 49 - 80 years). Our code is available at \url{https://github.com/cdttk/biv-volumetric-meshing/tree/plos2025}, and pre-trained networks, representative volumetric meshes with fibers and UVCs will be made available soon.

% % Please keep the Author Summary between 150 and 200 words
% % Use first person. PLOS ONE authors please skip this step. 
% % Author Summary not valid for PLOS ONE submissions.   
% \section*{Author summary}
% Lorem ipsum dolor sit amet, consectetur adipiscing elit. Curabitur eget porta erat. Morbi consectetur est vel gravida pretium. Suspendisse ut dui eu ante cursus gravida non sed sem. Nullam sapien tellus, commodo id velit id, eleifend volutpat quam. Phasellus mauris velit, dapibus finibus elementum vel, pulvinar non tellus. Nunc pellentesque pretium diam, quis maximus dolor faucibus id. Nunc convallis sodales ante, ut ullamcorper est egestas vitae. Nam sit amet enim ultrices, ultrices elit pulvinar, volutpat risus.

% Use "Eq" instead of "Equation" for equation citations.
\section*{Introduction}
Approximately one-third of deaths globally are estimated to be caused by cardiovascular diseases \cite{doi:10.1161/CIRCRESAHA.117.308903}. A Cardiac Digital Twin (CDT) aims to improve cardiac healthcare by creating a virtual replica of a patient's heart through an interdisciplinary approach in personalized medicine. The CDT can, in principle, be used to improve risk assessment, screening, diagnosis, and treatment by providing more accurate personalised in silico monitoring and predictions compared to traditional methods \cite{10.1093/eurheartj/ehaa159,Coorey2022,Viola2023}. A CDT is continuously updated with new relevant data - these data can be periodically acquired in hospital visits, including imaging studies, or can be acquired real-time, e.g. with wearable devices. A CDT-based monitoring and prediction system may involve a variety of components such as mechanistic simulation, statistical prediction models, many types of data sources, and secure data storage and transfer protocols. \cite{10.1093/eurheartj/ehaa159,Coorey2022,Viola2023}. 

In this context, CDTs based on recapitulating the anatomy and function of the ventricles of the heart are being used to predict sudden cardiac death risks \cite{Zaidi2024}, guide the placements of
implanted devices \cite{Qian2023}, plan ablation procedures \cite{Monaci2022}, better predict adverse cardiovascular events \cite{Mauger2023}, and provide mechanistic insights in public health by inferring patient-specific myocardial tissue properties \cite{Qian2023CV}, among others. A key requirement in all these applications is the ability to personalise a bi-ventricular cardiac computational mesh to each patient.

Cardiovascular Magnetic Resonance (CMR) images are routinely acquired in clinical practice for calculating diagnostic metrics such as left and right ventricular (LV and RV) volumes and ejection fraction (EF) \cite{RUIJSINK2020684}\cite{vonKnobelsdorff-Brenkenhoff2017}. CMR acquisitions typically consist of multiple 2D cine steady state free precision scans, leading to a set of slices in heart's short axis (SAX), usually 8-12 slices with a thickness around 8-10 mm, and one to three circumferentially sampled long axis (LAX) slices \cite{jimaging4010016,Banerjee2021}. Due to the sparse and cross-sectional nature of the slices, and the fact that the slices might be misaligned due to patient movement and differences in breath-hold position during acquisition, it is not straightforward to reconstruct the 3D structure from these 2D images \cite{jimaging4010016,Banerjee2021}. Hence, this problem has been an active research area and a variety of methods have been proposed.

One approach is to tackle the problem at the imaging protocol and reconstruction stage and attempt to create an isotropic 3D cine image through faster undersampled scans and smart image reconstruction using deep-learning models instead of classical reconstruction algorithms \cite{Kustner2020, Sandino2021, Hammernik2023}. While this approach shows promise for future applications, 2D cine is still the clinical standard and most existing cine MR datasets are acquired using the 2D clinical standard.

The methods proposed for 3D heart mesh reconstruction from typical clinical cine MR acquisitions have included three common approaches: deforming an initial/template mesh to fit the CMR data, fitting to a statistical shape model (SSM), and deep-learning based reconstruction. \cite{LAMATA2011801} warped a template mesh to fit the anatomy through registration of binary images created from the template mesh and segmentation of the medical image. \cite{Villard2018} and \cite{Banerjee2021} deformed an initial mesh to a point cloud obtained from contours of extracted relevant structures such as the LV endocardium and epicardium, and RV endocardium after segmentation. In \cite{mauger2019right}, a template biventricular heart mesh is deformed under diffeomorphic constraints to a point cloud consisting of contours and landmarks extracted from CMR images. \cite{Beetz2021} proposed a deep-learning approach where a point cloud completion network was trained to convert the sparse point cloud extracted from CMR images to a dense point cloud. In \cite{BanerjeeSSM2022}, a multi-step whole-heart mesh reconstruction method was proposed where the contours extracted from CMR images were first used to reconstruct a biventricular mesh, which was then fitted to a whole-heart SSM created from high-resolution CT by \cite{Hoogendoorn2013}. As a final step, the whole-heart mesh was deformed using the method proposed in \cite{Villard2018}. \cite{Xia2022} proposed a heart shape reconstruction method based on learning the principal component analysis parameters of a point distribution model. The training set was created by registering the cardiac atlas mesh from \cite{rodero2021linking} to manual contours extracted from CMR. This method simultaneously used the SAX view, 3 LAX views and patient metadata in the learning process. \cite{gaggion2023multiview} directly created a volumetric mesh from CMR images using a graph convolution network. The training set was created in the same way as \cite{Xia2022}. \cite{Kong2021,Kong2023} proposed two similar whole-heart mesh reconstruction methods based on learning mesh deformations using graph convolution networks coupled with a segmentation module. The methods were demonstrated to work with 3D CT, 3D MR and cine MR images. Although these methods are capable in principle of deriving digital twin representative models at scale, none have done so. The existing methods have one or more of the following limitations: either they did not explicitly model the locations of the aorta, mitral and tricuspid valves from available long axis images, or they were not demonstrated to work at scale (\textgreater50k cases), or source code is not publicly available. 

Reconstructing the 3D heart shape from CMR images requires considerable effort and expertise. Additionally, creating a full CDT system may be a complex interdisciplinary endeavour, and researchers that work on one particular part of the system may not be familiar with another part. Therefore, it is imperative that both open source tools and publicly available outputs are provided that can be readily used by researchers who work on other parts of the CDT system. This paper aims to facilitate cardiac digital twinning with the following contributions:

\begin{enumerate}
  \item A fully open source automatic pipeline from raw cine MR images to biventricular meshes. The method is validated on UKBB data and can both act as a baseline for other researchers who are working on the same problem, or can be used to easily create patient-specific meshes from cine MR images.
  \item 1423 publicly available representative meshes created from different sex, BMI, and age groupings. This diverse set of meshes can be readily used by other CDT researchers who use biventricular meshes as input to their work, such as researchers working on mechanistic simulation or statistical prediction models.
\end{enumerate}

\section*{Materials and methods}
\begin{figure}[!h]
\centering
\includegraphics[scale=.61]{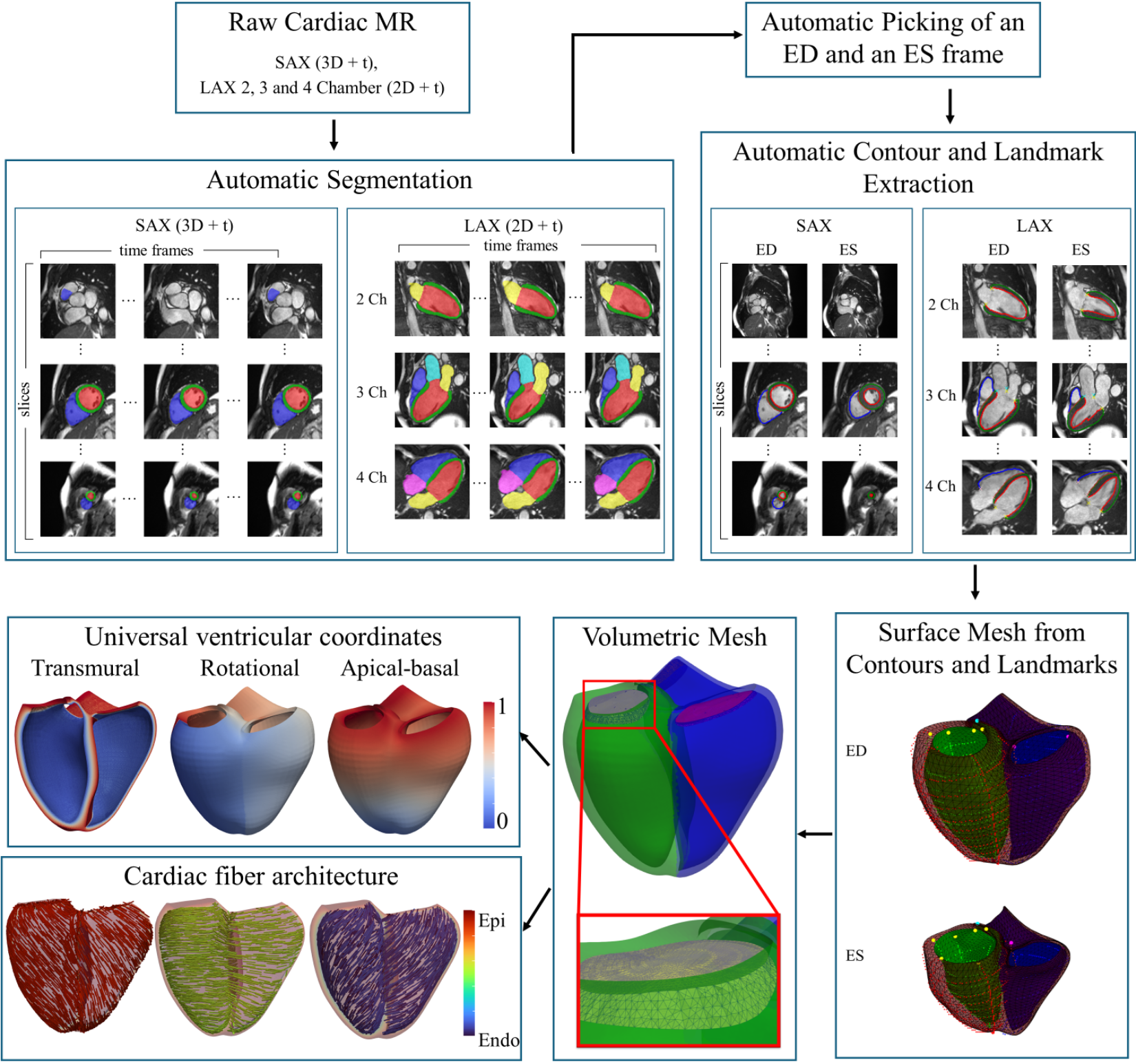}
\caption{{\bf Overview.}
An overview of the proposed pipeline. Reproduced by kind permission of UK Biobank \copyright.}
\label{overview}
\end{figure}

An overview of the proposed pipeline is illustrated in Fig~\ref{overview}. Briefly, raw SAX and LAX images were first automatically segmented. Using the segmentations, contours and landmarks were extracted for each view. A biventricular surface mesh was then reconstructed using the contours and landmarks. The surface mesh was then converted to a volumetric mesh and the volumetric mesh was used to reconstruct cardiac fiber architecture and also mapped to universal ventricular coordinates.

Legal and ethical approval for the study is covered by the UK Biobank's Research Tissue Bank approval (REC reference 11/NW/0382 for the initial approval and renewals with REC references 16/NW/0274 and 21/NW/0157) obtained from the North West Multi-centre Research Ethics Committee (MREC). Our research project was approved by the UK Biobank in accordance with their application procedure.

\subsection*{Dataset}
UK Biobank (UKBB) \cite{UKBB} is a large cohort study that includes CMR imaging and associated epidemiological and clinical data from \textgreater75,000 participants in the imaging substudy. The CMR imaging protocol has been described previously \cite{Petersen2016}. Briefly, steady-state free-precession CMR cine images were acquired in short and long axis locations, each with a separate breath-hold. Typical spatial resolution was 1.8x1.8x8mm for short axis and 1.6x1.6x6mm for long axis images. From this database, we utilized the following data for this study:
\begin{enumerate}
  \item Long-axis (LAX) CMR images. These include images from three different views: 2-chamber, 3-chamber, and 4-chamber. Each view is a 2D + t (time) image with fifty time frames. 
  \item Short-axis (SAX) CMR images. Each SAX image is a 3D + t image with fifty time frames.
  \item Age, sex and body mass index (BMI) of participants.
  \item Manual segmentations for the end-diastolic (ED) and end-systolic (ES) frames of the SAX view of 4788 participants. These manual segmentations were created in a previous study \cite{Petersen2016}. 
  \item Various derived phenotypes such as ventricle volumes and ejection fractions reported in two previous studies for validation and comparison. \cite{Petersen2016, bai2020population}.
\end{enumerate}
CMR images and SAX manual segmentations were downloaded from UKBB on March 2023 and other data were downloaded from UKBB on March 2024. For more details on the dataset, see Table~\ref{S1_Table}.

\subsection*{Preprocessing}
The only preprocessing we performed was the conversion of raw DICOM files from UKBB corresponding to the LAX and SAX images, into NIFTI images. For each participant, four NIFTI files were created that corresponded to the four different CMR views: LAX 2-chamber, LAX 3-chamber, LAX 4-chamber, and SAX.

\subsection*{Segmentation}
For each CMR view, a different set of heart structures were selected for segmentation as follows:
\begin{enumerate}
  \item LAX 2-chamber: LV cavity, LV myocardium, LA cavity.
  \item LAX 3-chamber: LV cavity, LV myocardium, RV cavity, RA cavity, Aorta
  \item LAX 4-chamber: LV cavity, LV myocardium, RV cavity, LA cavity, RA cavity
  \item SAX: LV cavity, LV myocardium, RV cavity
\end{enumerate}
The structures to segment were selected based on how reliably they can be segmented from each view. RV myocardium, for example, is typically excluded from cine MR segmentation because it is too thin and usually gets lost to partial volume effect at typical cine MR resolution.

For the SAX view, we utilized manual segmentations that were previously created for the ED and ES frames in a previous study \cite{Petersen2016}. The identifier of 4788 participants could be matched between the manual segmentations and raw CMRs and these were used in this study. For the LAX views, we manually segmented 150 participants at the ED and ES frames selected in no particular order from the same set of 4788 participants. The segmentations were verified and corrected under the guidance of a CMR level 3 clinical practitioner. 4000 of the participants with manual SAX segmentations, and 100 of the participants with manual LAX segmentations were used for supervised training of segmentation networks. The remaining participants, 788 for SAX and 50 for LAX, were used as the test sets for evaluation. Training and test subjects were selected randomly.

For automatic segmentation, there are a wide variety of methods proposed in literature (see e.g. \cite{ChenChen2020, Ammar2021, Tayebi2023, ElTaraboulsi2023, Alnasser2024}). For this study, we picked the nnUNet \cite{Isensee2021} framework which implements self-configuring UNet-based \cite{UNet} networks for medical image segmentation, and is reported to achieve state-of-the-art performance in various recent medical segmentation challenges \cite{mm,ji2022amos,mm_rv}.
The recommended usage of training five folds from the training set, and using all of them during inference was deployed separately for all views. Only 2D training was used for all views. Since LAX views are 2D, that is the only option, and for SAX, we chose to use only 2D because there was no clear evidence of 3D being better in a previous study that used nnUNet for CMR segmentation \cite{Full2021}. The ED and ES frames were treated as independent images for segmentation.

\subsection*{Automatic selection of ED and ES frames}
\begin{figure}[!h]
\centering
\includegraphics[scale=.60]{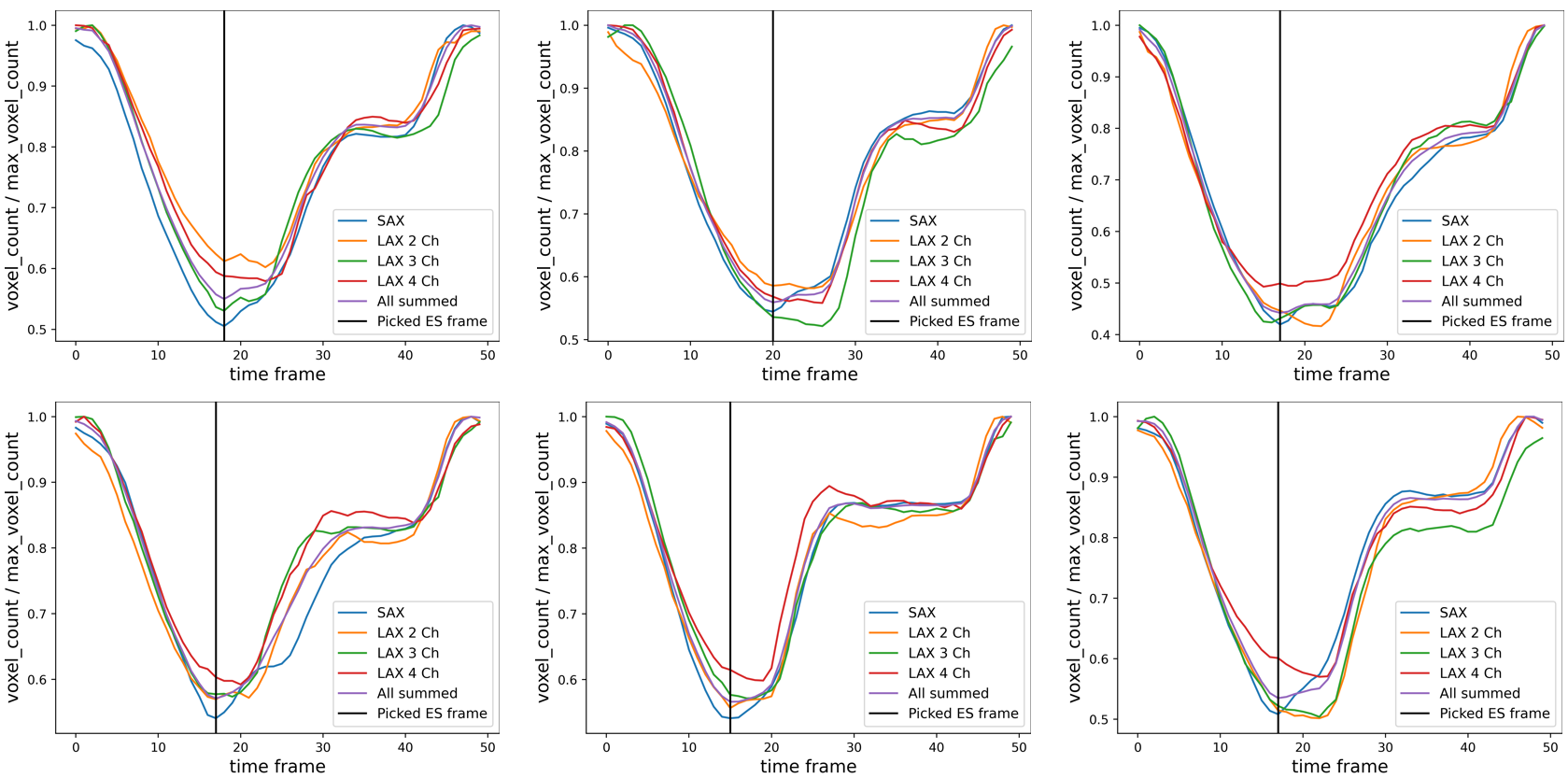}
\caption{{\bf Picking the ES frame.}
The LV volume transients (LV voxel count on each time frame divided by the maximum LV voxel count across time frames) of six example participants for all the views separately, and their sum across the views. For the SAX view, only the five mid-slices were used for the calculation. The ES frame was picked as the time frame that minimizes the sum across all views.}
\label{es_picking}
\end{figure}

The evaluation of segmentation accuracy, and subsequent meshing steps were performed for only the ED and ES frames because manual segmentations and previously reported reference derived phenotypes based on manual segmentations were only available for ED and ES frames, which makes validation and evaluation difficult for other frames. Hence, although the proposed pipeline could be applied to all time frames in principle, ED and ES frames were automatically selected after segmentation before subsequent steps. For the ED frame, we simply selected the first time frame since that is the first frame acquired after the R wave detection in the UKBB CMR protocol, as done in \cite{Petersen2016}. For the ES frame, we picked the time frame for which the sum of voxel counts labelled as ``LV cavity" in the LAX views, and the five mid-slices of the SAX view, was minimum. Automatic ES selection is illustrated for six example participants in Fig~\ref{es_picking}. The reason we did not use a similar algorithm for picking the ED frame as the frame with the maximum LV cavity volume is that, this algorithm is not perfect and can pick the wrong frame if the automatic segmentation fails in some of the time frames. For the UKBB dataset specifically, we think picking the first frame as the ED is more robust due to the UKBB imaging protocol. On other datasets where this is not the case however, picking the maximum LV cavity volume is a good alternative.

\subsection*{Extraction of contours and landmarks from the segmentation masks}
After the ED and ES frames are selected, we then extracted the following contours and landmarks from each of the four CMR views using their segmentation masks:
\begin{enumerate}
  \item LAX 2 chamber: Contours: LV endocardial, LV epicardial. Landmarks: Two points on the mitral valve plane, and the apex point, which is defined as the furthest point on the LV epicardial contour, from the mid-point between the two selected points on the mitral valve plane.
  \item LAX 3 chamber: Contours: LV endocardial, LV epicardial, RV septum, RV free wall. Landmarks: Two points on the aorta-LV intersection, and two points on the mitral valve plane. 
  \item LAX 4 chamber: Contours: LV endocardial, LV epicardial, RV septum, RV free wall. Landmarks: Two points on the mitral valve plane, and two points on the tricuspid valve plane.
  \item SAX: Contours: LV endocardial, LV epicardial, RV septum, RV free wall.
\end{enumerate}

\begin{figure}[!h]
\centering
\includegraphics[scale=.28]{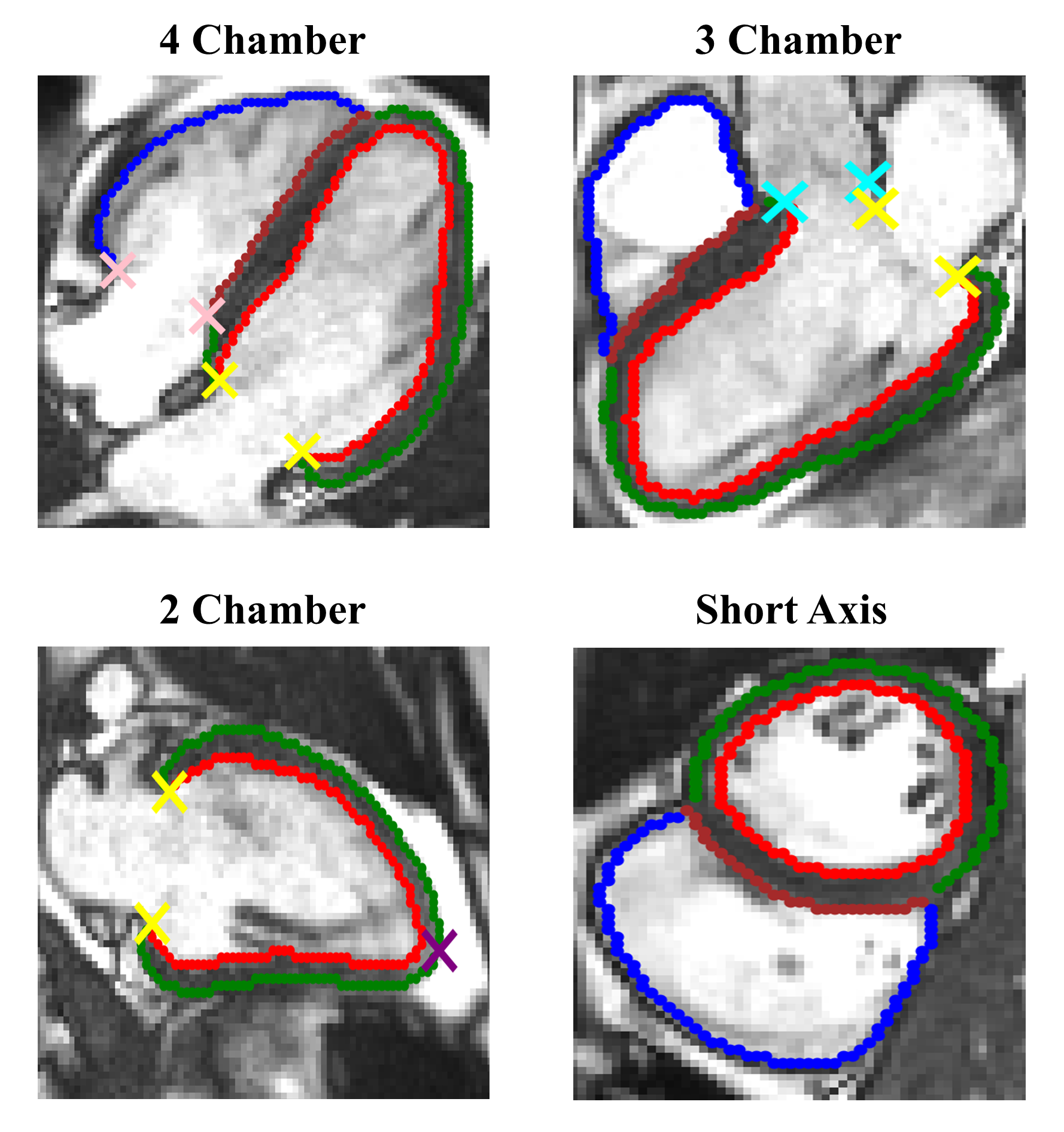}
\caption{{\bf Contour and landmark extraction from segmentations.}
Illustration of contours and landmarks automatically extracted from the ED frame of a randomly selected participant. Contours: Red: LV endocardium, Green: LV epicardium, Brown: RV septum, Blue: RV free wall. Landmarks: Yellow: Points on the mitral valve, Pink: Points on the tricuspid valve, Teal: Points on the aorta-LV intersection, Purple: Apex. Reproduced by kind permission of UK Biobank \copyright.}
\label{contour_landmark}
\end{figure}

The contours and landmarks automatically extracted from the ED frame of a random participant are illustrated in Fig~\ref{contour_landmark}.

The extraction algorithm is based on simple heuristics, and the reader is referred to the source code and documentation for details.

\subsection*{Finite element mesh construction}
We used an established Atlas-based pipeline\cite{mauger2019right} to construct the biventricular surface meshes. As the RV myocardium was not captured in the segmentation, during the surface mesh generation, the RV epicardium was estimated by extending the RV endocardium points normal to the surfaces by 3 mm consistent with experimental measurements\cite{matsukubo1977,ho2006anatomy}. The resultant surface meshes were then used to construct tetrahedral finite element meshes using Meshtool\cite{neic2020automating} including six distinct regions of LV myocardium, RV myocardium, aortic, tricuspid, pulmonary and mitral valves. 

All volumetric meshes were then incorporated with a morphological coordinate system, which describes the positions within ventricles based on the apical-basal (Z), transmural ($\rho$) (from endocardium to epicardium), rotational ($\Phi$) (anterior, anteroseptal, inferior, inferolateral, anterolateral) and chamber-wise (left ventricle and right ventricle) coordinates \cite{Bayer2018}. For each mesh, it also includes a realistic biventricular myocardial fibre structure implemented using a rule-based approach with a transmural variation of angle $\alpha$ as from 60\degree\ to -60\degree\ in longitudinal fibre directions and angle $\beta$ as from -65\degree\ to 25\degree\ in transverse fibre directions from endocardium to epicardium \cite{Bayer2012}.

\subsection*{Representative mesh generation for different sex, age and BMI groups}
Before constructing representative meshes, we employed a mesh quality control step to remove meshes that showed a large difference between surface-derived and segmentation-derived phenotypes from the SAX view. Specifically, LV and RV volumes, and LV myocardium mass were derived directly from nnUNet segmentations of the SAX view, and from surface meshes. The LV mass was computed from the end-diastolic LV myocardial volume using a density of 1.05 g/mL. The relative differences were then calculated between the phenotype derived directly from SAX segmentations and derived from surface meshes for each phenotype for the whole cohort. Finally, the participants for which the relative difference for at least one phenotype was greater than 75th percentile plus 1.5 times the interquartile range were not included in the construction of representative meshes.

Participants with viable surface meshes were categorized based on their demographics of sex, age and BMI. The age bins start with 44 with incremental of 1 year for each bin (i.e. 44, 45, 46…85) while the BMI bins start with 15 kg/m$^2$ with incremental of 1 kg/m$^2$ for each bin (i.e. 15,16,17,…,50). For each unique sex, age, BMI group with at least three participants, the ‘average’ surface mesh was constructed. Before averaging, all surface meshes were first registered to a common reference space to remove the inter-participant variability in orientations and positions using a Procrustes-based method. Specifically, we first randomly selected a participant's surface mesh as the reference and then superimposed all other surface meshes to this reference. To avoid the bias towards this reference, we computed the mean mesh of the set of superimposed meshes. Then the Procrustes distances  between the mean mesh and the reference mesh were computed and if higher than 1e-6, the reference was updated as the mean mesh until the optimal reference mesh is found. In this way, we found the optimal reference mesh which represents the average heart position and orientation. Finally, all surface meshes were aligned with the optimal reference mesh and the average volumetric meshes for different bins were constructed along with their UVCs and ventricular fibres.

% Results and Discussion can be combined.
\section*{Results}
\subsection*{Segmentation validation}
The segmentation networks were validated in two ways: Comparison to manual segmentations using Dice scores, and comparison to previously reported derived phenotypes in literature such as volumes and ejection fraction. The participants in the test sets, 788 for SAX and 50 for LAX, were utilized for both of these comparisons.

Mean Dice scores for the test sets of each view for each segmented structure are given in Table~\ref{seg_dice}. The automatic segmentations have high volumetric overlap with manual segmentations for every view and segmented structure.

\begin{table}[!ht]
\centering
\caption{
{\bf Dice scores: Manual vs. nnUNet segmentations.}}
\label{seg_dice}
\begin{tabular}{|l|l|l|l|l|}
\hline
    & \bf{2Ch}     & \bf{3Ch}       & \bf{4Ch}       & \bf{SAX}       \\ \thickhline
  LV ED & 0.97 (0.01) & 0.98 (0.01) & 0.97 (0.01) & 0.97 (0.02)\\ \hline
  LV ES & 0.88 (0.10) & 0.94 (0.06) & 0.89 (0.09) & 0.93 (0.04)\\ \hline
  RV ED &  & 0.95 (0.03) & 0.95 (0.02) & 0.93 (0.04)\\ \hline
  RV ES &  & 0.88 (0.08) & 0.86 (0.10) & 0.88 (0.05)\\ \hline
  Myo ED & 0.88 (0.02) & 0.91 (0.03) & 0.89 (0.03) & 0.88 (0.03)\\ \hline
  Myo ES & 0.86 (0.09) & 0.93 (0.07) & 0.86 (0.09) & 0.90 (0.03)\\ \hline
  LA ED & 0.90 (0.06) & 0.96 (0.06) & 0.89 (0.06) & \\ \hline
  LA ES & 0.95 (0.03) & 0.98 (0.03) & 0.95 (0.03) & \\ \hline
  RA ED &  &  & 0.92 (0.06) & \\ \hline
  RA ES &  &  & 0.96 (0.04) & \\ \hline
  Ao ED &  & 0.97 (0.02) &  & \\ \hline
  Ao ES &  & 0.96 (0.05) &  & \\ \hline
\end{tabular}
\begin{flushleft} Mean and standard deviations of Dice scores between manual and nnUNet-produced segmentations. Empty cells mean that the structure is not segmented from the corresponding view.
\end{flushleft}
\end{table}

\subsection*{Derived phenotypes}
LV and RV volumes, and LV myocardium mass derived directly from manual, nnUNet-based and UNet-based (reported in \cite{BaiWenjia2018}) SAX segmentations, and from surface meshes are compared to each other in Table~\ref{sax_derived_measures}. The number of participants used for this table was 731, which is smaller than the SAX test set size of 788 because it is also required that a UNet-based result exists for all the phenotypes and the surface reconstruction succeeds.

\begin{table}[!ht]
\begin{adjustwidth}{-2.25in}{0in}
\centering
\caption{\bf{Phenotypes derived from the SAX view.}}\label{sax_derived_measures}
\begin{tabular}{|l|l|l|l|l|}
\hline
      \bf{(a) Value} & \bf{Manual} & \bf{nnUNet} & \bf{UNet (Bai)} & \bf{Mesh (Ours)} \\ \thickhline
  LVEDV (mL) & 150.6 (34.2) & 149.8 (33.4) & 150.9 (33.3) & 139.3 (30.8)\\ \hline
  LVESV (mL) & 63.1 (19.1) & 62.3 (18.5) & 62.4 (18.6) & 59.6 (17.3)\\ \hline
  LVM at ED (gram)& 92.4 (24.4) & 94.1 (23.2) & 86.9 (21.8) & 112.8 (25.4)\\ \hline
  RVEDV (mL) & 157.6 (38.7) & 157.0 (37.6) & 158.6 (38.0) & 145.5 (34.9)\\ \hline
  RVESV (mL) & 70.4 (23.0) & 68.9 (21.1) & 69.4 (21.9) & 64.0 (19.6)\\ \hline
  LV EF (\%) & 58.4 (6.1) & 58.7 (5.8) & 59.0 (5.7) & 57.6 (5.3)\\ \hline
  RV EF (\%) & 55.9 (6.2) & 56.5 (5.7) & 56.7 (5.9) & 56.5 (5.6)\\ \thickhline
    \bf{(b) Absolute diff.} & \bf{nnUNet-Manual} & \bf{UNet-Manual} & \bf{nnUNet-UNet} & \bf{nnUNet-Mesh} \\ \thickhline
  LVEDV (mL) & 6.0 (4.7) & 5.9 (4.5) & 2.5 (2.6) & 11.1 (6.9)\\ \hline
  LVESV (mL) & 5.4 (4.8) & 5.1 (4.6) & 2.3 (2.9) & 4.9 (3.9)\\ \hline
  LVM at ED (gram) & 7.1 (5.5) & 8.2 (6.2) & 7.2 (2.9) & 18.8 (5.2)\\ \hline
  RVEDV (mL) & 7.9 (7.4) & 8.2 (7.4) & 4.4 (4.1) & 12.0 (7.3)\\ \hline
  RVESV (mL) & 6.9 (6.8) & 6.6 (6.4) & 2.8 (2.6) & 5.6 (4.1)\\ \hline
  LV EF (\%) & 3.1 (2.8) & 3.0 (2.8) & 1.5 (2.0) & 3.1 (2.7)\\ \hline
  RV EF (\%) & 3.9 (3.3) & 3.9 (3.1) & 1.9 (1.8) & 2.3 (2.2)\\ \thickhline
      \bf{(c) Relative diff.} & \bf{nnUNet-Manual} & \bf{UNet-Manual} & \bf{nnUNet-UNet} & \bf{nnUNet-Mesh} \\ \thickhline 
  LVEDV (\%) & 4.0 (3.0) & 4.0 (2.9) & 1.7 (1.8) & 7.7 (4.7)\\ \hline
  LVESV (\%) & 8.8 (7.5) & 8.3 (7.3) & 3.7 (4.6) & 8.0 (5.7)\\ \hline
  LVM at ED (\%) & 7.9 (6.3) & 9.1 (6.5) & 8.0 (2.7) & 18.6 (4.8)\\ \hline
  RVEDV (\%) & 5.0 (4.2) & 5.1 (4.2) & 2.7 (2.5) & 8.0 (4.7)\\ \hline
  RVESV (\%) & 9.9 (8.7) & 9.6 (8.5) & 4.1 (3.7) & 8.6 (5.6)\\ \hline
\end{tabular}
\begin{flushleft} Mean and standard deviations of the phenotypes derived from the SAX view with different methods, and their absolute and relative differences (n=731).
\end{flushleft}
\end{adjustwidth}
\end{table}

LA and RA volumes derived from from manual, nnUNet-based and UNet-based (reported in \cite{BaiWenjia2018}) LAX segmentations are reported in Table~\ref{lax_derived_measures}. Since the meshes are biventricular, and hence do not include the atria, these cannot be derived from the meshes. The number of participants used for this table was 48, which is smaller than the LAX test set size of 50 because it is also required that a UNet-based result exists for all the phenotypes.

\begin{table}[!ht]
\centering
\caption{\bf{Phenotypes derived from the LAX views.}}\label{lax_derived_measures}
\begin{tabular}{|l|l|l|l|}
\hline
  \bf{(a) Value} & \bf{Manual (Petersen)} & \bf{nnUNet} & \bf{UNet(Bai)} \\ \thickhline
  LAMinV (mL) & 29.0 (11.1) & 31.0 (11.0) & 30.5 (11.4)\\ \hline
  LAMaxV (mL) & 70.9 (21.3) & 77.3 (22.1) & 76.5 (21.8)\\ \hline
  RAMinV (mL) & 48.3 (20.8) & 47.2 (17.2) & 47.8 (18.8)\\ \hline
  RAMaxV (mL) & 79.7 (24.8) & 84.9 (24.2) & 86.9 (27.5)\\ \thickhline
  \bf{(b) Absolute diff.} & \bf{nnUNet-Manual} & \bf{UNet-Manual} & \bf{nnUNet-UNet} \\ \thickhline
  LAMinV (mL) & 3.5 (2.7) & 3.0 (2.4) & 2.5 (2.2)\\ \hline
  LAMaxV (mL) & 8.1 (5.4) & 6.6 (3.8) & 4.4 (4.0)\\ \hline
  RAMinV (mL) & 4.7 (3.7) & 3.9 (3.4) & 2.9 (2.4)\\ \hline
  RAMaxV (mL) & 7.2 (5.0) & 7.8 (6.3) & 4.5 (4.6)\\ \thickhline
  \bf{(c) Relative diff.} & \bf{nnUNet-Manual} & \bf{UNet-Manual} & \bf{nnUNet-UNet} \\ \thickhline
  LAMinV (\%) & 12.5 (8.7) & 10.9 (8.0) & 9.0 (7.9)\\ \hline
  LAMaxV (\%) & 11.3 (7.9) & 9.2 (5.1) & 5.8 (5.4)\\ \hline
  RAMinV (\%) & 10.1 (6.9) & 8.4 (6.7) & 6.5 (6.0)\\ \hline
  RAMaxV (\%) & 9.0 (6.5) & 9.4 (5.8) & 4.9 (3.8)\\ \hline
\end{tabular}
\begin{flushleft} Mean and standard deviations of the phenotypes derived from the LAX views with different methods, and their absolute and relative differences (n=48).
\end{flushleft}
\end{table}

\subsection*{Application: Representative meshes}
\begin{figure}[!h]
\centering
\includegraphics[scale=13.3]{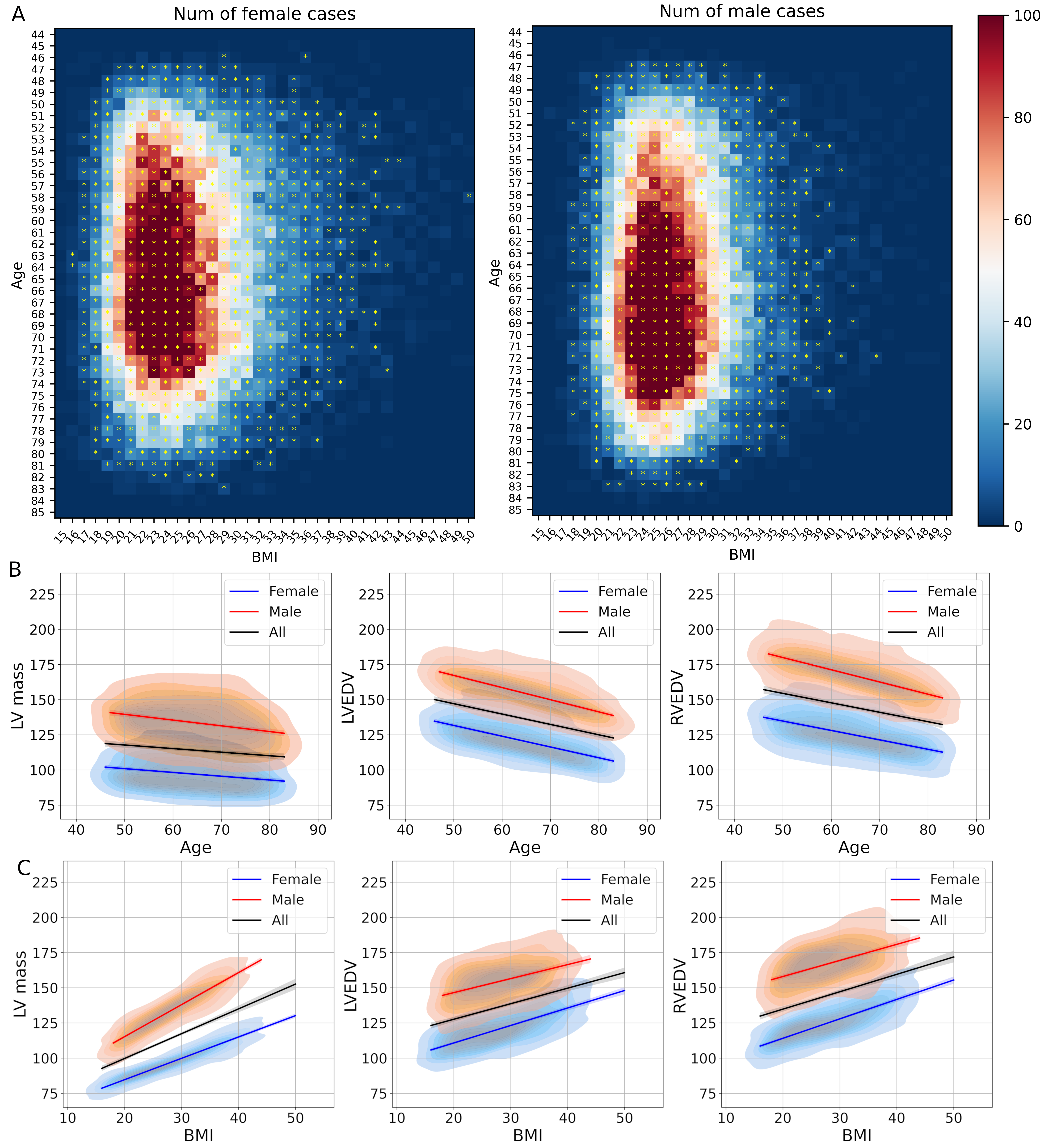}
\caption{{\bf Demographic bins and associations with derived phenotypes.}
(A) The number of female and male participants in this cohort in specific age and BMI bins. The bins that contain more than three participants are marked with a * and a representative mesh is created for each of these bins. The colour bar represents the absolute value of the number of participants. (B),(C) Associations of derived phenotypes of all representative hearts with sex, age and BMI. The derived phenotypes LVEDV, RVEDV and LV mass are plotted as kernel density plots along with linear-regression lines for the whole cohort (black), for female (blue) and for male (red) (n=1423).}
\label{bin_heatmaps}
\end{figure}

\begin{figure}[!h]
\centering
\includegraphics[scale=12.6]{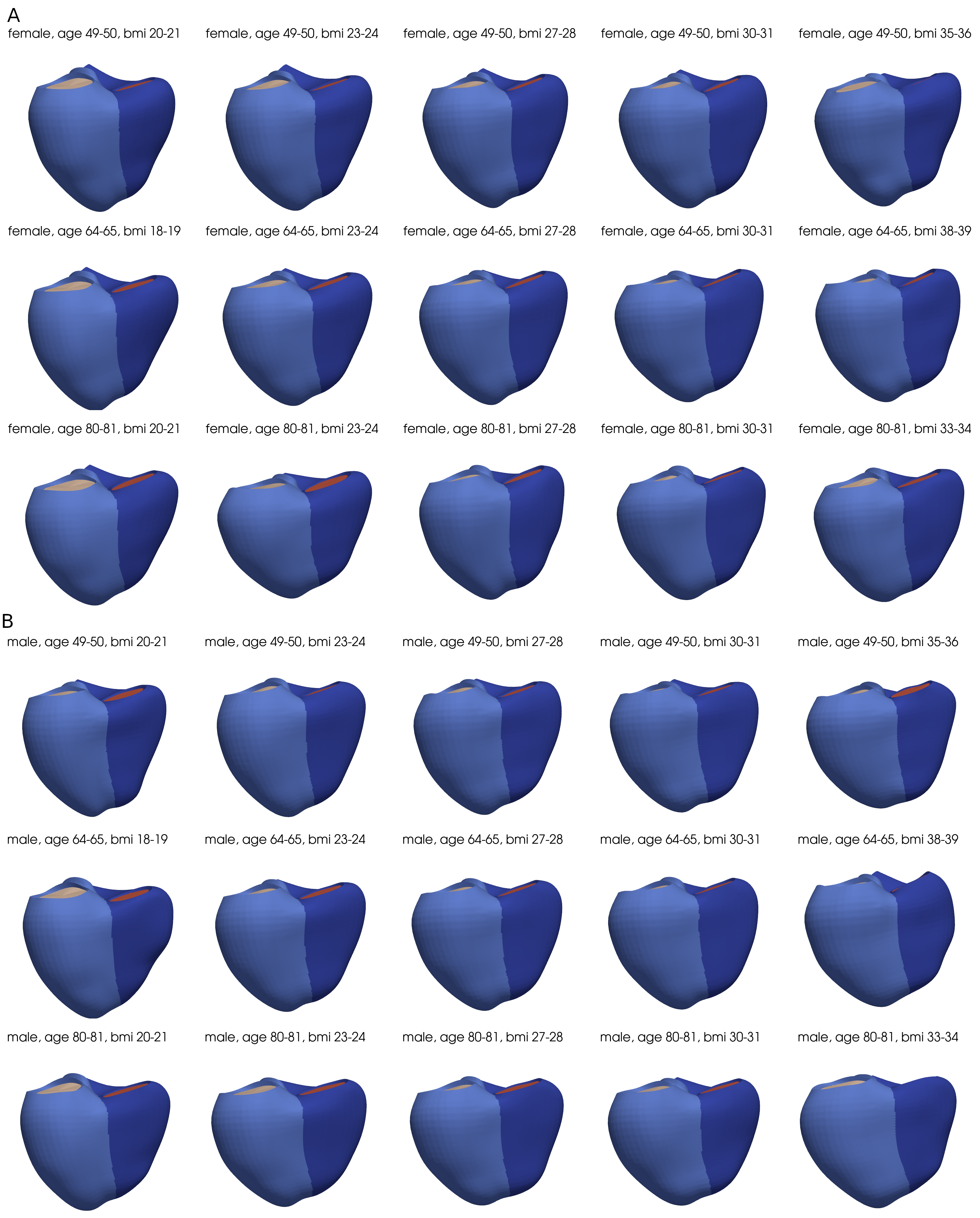}
\caption{{\bf Examples of representative meshes for different demographic bins.}}
\label{mesh_visual}
\end{figure}

Surface meshes for the whole cohort were constructed from the end diastolic frames and used to build a cohort of representative meshes for specific gender, age and BMI bins. Fig~\ref{bin_heatmaps} (A) shows the number of females and males in the whole cohort, categorising into specific age (44 to 85 years old) and BMI (15 to 50 kg/m$^2$) bins. The bins that contained three or more participants were used to construct a representative mesh for that bin, resulting in a total number of 1423 representative meshes computed from 46917 individual meshes. Some example representative meshes are illustrated in Fig~\ref{mesh_visual}.

For details on the number of participants and meshes utilized at each step of the pipeline, see Table~\ref{S2_Table}.

We investigated the changes of cardiac structures with sex, BMI and age in the cohort of representative meshes. We computed three phenotypes: LVEDV, RVEDV and LV mass, and fitted a linear regression model for females and males with different age and BMI as shown in Fig~\ref{bin_heatmaps} (B) and (C). We found that all three phenotypes are greater for males than for females (LV mass: $133.6\pm14.8$ vs $97.2\pm11.3$ g; LVEDV: $154.5\pm12.9$ vs $121.0\pm12.5$ mL; RVEDV: $167.2\pm13.5$ vs $125.5\pm12.8$ mL, all $P<0.0001$ from Mann-Whitney U test). Aging was associated with a reduction in both LV and RV diastolic volumes and LV mass ($-0.8$ and $-0.7$ mL/year and $-0.3$ g/year, $\beta=184,188$ and $130.5$, $P=5\times10^{-38}$, $4.6\times10^{-23}$ and $4.6\times10^{-5}$). In contrast, increased BMI leads to an increase in LV and RV diastolic volumes and LV mass (1.1 and 1.2 mL/(kg/m$^2$) and 1.8 g/(kg/m$^2$), $\beta=$105.5, 110.1 and 64.5, P=$9.1\times10^{-39}$, $2.8\times10^{-35}$ and $7.9\times10^{-92}$).

\section*{Discussion}
In this paper, we presented an automatic pipeline for CDT construction from raw cine MR images to biventricular meshes and validated it on the CMR data of 54926 participants from the UKBB. We also computed 1423 representative meshes and their corresponding fibers and UVCs from different sex, BMI, and age groupings from the UKBB. All the code used in the study, the trained segmentation networks, the representative meshes, and their corresponding fibers and UVCs are made publicly available. This will facilitate cardiac digital twinning from CMR images by providing an easily reproducible baseline for other researchers and readily usable meshes for mechanistic simulations and statistical prediction models.

\subsection*{Segmentation}
The accuracy of the trained segmentation networks were validated using Dice scores and derived phenotypes. The Dice scores showed very good agreement between manual and automatic segmentations for all CMR views and structures. Due to the large time cost of manual segmentation, an inter-observer variability study was not conducted. In literature, \cite{BaiWenjia2018} previously reported inter-observer variability on a set of SAX images of 50 subjects using three observers. Mean inter-observer Dice scores were 0.93 for LV, 0.88 for LV myocardium, and 0.88 for RV. The ED and ES scores were not reported separately, but it can be seen that our nnUNet-manual Dice scores are slightly higher if we average the ED and ES dice scores in Table \ref{seg_dice} (0.95 for LV, 0.89 for LV myocardium, and 0.91 for RV).

It is worth noting that structures that are larger in the ED frame compared to the ES frame, such as the left and right ventricle cavities, tend to have a higher Dice score in the ED frame, and similarly, the atria which are smaller in the ED frame have a higher Dice score in the ES frame. This is not surprising since a volumetric overlap measure like Dice score tends to give higher scores for large connected structures due to the non-overlapping regions occurring only near the edges of the structure. For the LV myocardium, this effect is not clearly observed, which is likely due to the greater inherent ambiguity in the LV myocardium segmentation on the ES frame that causes a Dice score-reducing effect.

\subsection*{Derived phenotypes}
The phenotypes derived from automatic vs. manual segmentations of the short-axis view showed similar agreement to the method presented in \cite{BaiWenjia2018}, which was reported to be comparable to inter-observer differences. This is not surprising since both methods are based on the U-Net architecture and trained using the manual segmentations from \cite{Petersen2016}. For the phenotypes derived from the long-axis, our method showed slightly higher difference to manual segmentations from \cite{Petersen2016} compared to the method from \cite{BaiWenjia2018}, but this is likely due to the fact that we trained our LAX networks using our own manual segmentations whereas \cite{BaiWenjia2018}'s method was trained on \cite{Petersen2016}'s LAX segmentations.

As seen in Table \ref{sax_derived_measures}, the phenotypes computed from the surface meshes showed a systematic bias in volumes and LV myocardial mass compared to segmentation-derived phenotypes from the nnUNet segmentations that were used to create the surface meshes. The mean of the LV-ED, LV-ES, RV-ED and RV-ES volumes computed from the meshes were respectively 7.0\%, 4.3\%, 7.3\% and 7.1\% smaller than the volumes computed from the nnUNet segmentations, and the mean of the LV myocardial mass was 19.9\% larger. An underestimation of volumes and overestimation of the LV myocardial mass derived from meshes was previously reported in \cite{Maciej2021} with a smaller difference (4.1\% smaller LV-ED volume and 6.6\% larger LV myocardial mass). One explanation of the difference is in the volume integration technique: the addition of disks from a short axis stack is different than the actual integration of volume from a 3D computational mesh. We would expect differences to be greater at the basal region of the anatomy. It is unclear at this stage how different meshing methods might affect this discrepancy and whether segmentation-derived or mesh-derived phenotypes should be preferred for a certain clinical application. Further study is required to address these questions. Interestingly, since the proportional volume underestimation is similar in the ED and ES frames, the difference in the LV and RV ejection fractions (important measures in clinical cardiology) remain comparable to the inter-observer variability reported in \cite{BaiWenjia2018}.

\subsection*{Representative meshes}
The cohort of representative meshes was constructed by averaging the hearts of participants falling into same age, sex and BMI bins. The LV and RV end-diastolic volumes and LV mass derived from the representative meshes were found to decrease with aging and increase with larger BMI, which is consistent with previous studies measured from the general population of the UK Biobank \cite{bai2020population,mauger2019right} as well as other independent datasets \cite{wade2018assessing}. Compared to the previously published virtual heart cohorts\cite{Qian2023,rodero2021linking,strocchi2020publicly}, this presented cohort (n=1423) is, to the best of our best knowledge, the largest heart cohort in the world which is additionally augmented with detailed demographics information such as sex, age and BMI, that can enable next generation population-specific studies. This cohort of anatomical-detailed representative models offers enriched heart shape information which provides a normative reference framework  that can complement and strengthen the previous studies on biological age estimation \cite{Mao2024,RaisiEstabragh2022,Ecker2023}, by enabling precise estimation of heart age deviation, supporting biological age acceleration analysis, improving prediction precision beyond radiomic approaches, and improving individualized cardiovascular risk profiling.

\subsection*{Limitations}
Since the proposed method relies on segmentation networks trained with supervised learning, domain shift issues could negatively affect segmentation performance on different datasets. While the UKBB is a large dataset, all the participants are volunteers, and hence are likely to be healthier than patient cohorts from hospitals. Further, all UKBB CMR images are obtained using the same imaging protocol and scanner model. Therefore, when using a different dataset, domain shift is likely with respect to pathology composition of the cohort, imaging protocol, and scanner model. Possible domain shift issues are not addressed in this paper and will require further detailed studies.

As the RV wall thickness is difficult to estimate from only CMR images, we assumed a fixed 3 mm as the RV wall thickness, that can lead to inaccurate RV function estimation. We set the RV wall thickness as a variable in our pipeline which can be easily adjusted to allow flexible usage for any new studies if a personalized estimation of RV wall thickness is available.

The meshing method can produce intersections between the RV septum and the RV free wall surfaces, which causes failures in conversion from surface to volume meshes. Thus, some meshes may require post-processing before being used in simulations.

The proposed pipeline processes each time frame separately and temporal smoothness is not guaranteed or analysed. Hence, the method may not be readily suitable for functional twinning of the heart motion.

\section*{Conclusion}

We presented an automatic pipeline from raw cine MR images to biventricular meshes and made publicly available all the code, trained segmentation networks and 1423 representative meshes and their corresponding fibers and UVCs from different sex, BMI, and age groupings from the UKBB. Future work is to create new representative cohorts for specific patient populations using the proposed pipeline, based on available individual data of summary diagnosis in the UKBB, enabling disease-specific studies. We anticipate this to be a valuable resource for other researchers working on cardiac digital twinning.

\section*{Acknowledgments}
This research has been conducted using the UK Biobank Resource under application number 88878.

\nolinenumbers

\clearpage

\section*{Supporting information}

\makeatletter
\renewcommand \thesection{S\@arabic\c@section}
\renewcommand \thetable{S\@arabic\c@table}
\renewcommand \thefigure{S\@arabic\c@figure}
\makeatother
\setcounter{figure}{0}
\setcounter{table}{0}

\begin{table}[!ht]
\begin{adjustwidth}{-2.25in}{0in} % Comment out/remove adjustwidth environment if table fits in text column.
\centering
\caption{
{\bf The dataset utilized from the UKBB.}}
\begin{tabular}{|p{0.2\linewidth} | p{0.8\linewidth}|}
\hline
\bf{UKBB field id} & \bf{Description}\\ \thickhline
31-0.0 & Sex of participant\\ \hline
34-0.0 & Year of birth of participant\\ \hline
52-0.0 & Calendar month of birth of participant\\ \hline
53-2.0 & Date of the first imaging visit\\ \hline
20208-2.0 & Long axis heart images acquired at the first imaging visit \\ \hline
20209-2.0 & Short axis heart images acquired at the first imaging visit \\ \hline
21001-2.0 & Body mass index of participant at the first imaging visit\\ \hline
24100-2.0 & LV end diastolic volume at the first imaging visit. See \cite{BaiWenjia2018,bai2020population} for methodology. \\ \hline
24101-2.0 & LV end systolic volume at the first imaging visit. See \cite{BaiWenjia2018,bai2020population} for methodology. \\ \hline
24103-2.0 & LV ejection fraction at the first imaging visit. See \cite{BaiWenjia2018,bai2020population} for methodology. \\ \hline
24105-2.0 & LV myocardial mass at the first imaging visit. See \cite{BaiWenjia2018,bai2020population} for methodology. \\ \hline
24106-2.0 & RV end diastolic volume at the first imaging visit. See \cite{BaiWenjia2018,bai2020population} for methodology. \\ \hline
24107-2.0 & RV end systolic volume at the first imaging visit. See \cite{BaiWenjia2018,bai2020population} for methodology. \\ \hline
24109-2.0 & RV ejection fraction at the first imaging visit. See \cite{BaiWenjia2018,bai2020population} for methodology. \\ \hline
24110-2.0 & LA maximum volume at the first imaging visit. See \cite{BaiWenjia2018,bai2020population} for methodology. \\ \hline
24111-2.0 & LA minimum volume at the first imaging visit. See \cite{BaiWenjia2018,bai2020population} for methodology. \\ \hline
24114-2.0 & RA maximum volume at the first imaging visit. See \cite{BaiWenjia2018,bai2020population} for methodology. \\ \hline
24115-2.0 & RA minimum volume at the first imaging visit. See \cite{BaiWenjia2018,bai2020population} for methodology. \\ \hline
31065-2.0 & RA maximum volume at the first imaging visit. See \cite{Petersen2016} for methodology. \\ \hline
31066-2.0 & RA minimum volume at the first imaging visit. See \cite{Petersen2016} for methodology. \\ \hline
31075-2.0 & LA maximum volume at the first imaging visit. See \cite{Petersen2016} for methodology. \\ \hline
31076-2.0 & LA minimum volume at the first imaging visit. See \cite{Petersen2016} for methodology. \\ \hline
Return 2541 &  This return from study \cite{Petersen2016} contains various data derived from heart MRI, we used the manual segmentation files after converting from cvi42 format to segmentation masks. \\ \hline
\end{tabular}
\label{S1_Table}
\end{adjustwidth}
\end{table}

\begin{table}[!ht]
\centering
\caption{
{\bf Number of subjects or meshes utilized at different steps of the pipeline.}}
\begin{tabular}{|p{0.85\linewidth} | p{0.15\linewidth}|}
\hline
\bf{Description} & \bf{\#}\\ \thickhline
Initial number of subjects that have both LAX and SAX DICOMs & 55835 \\ \hline
Number of subjects that have all four views after DICOM to NIFTI conversion & 54926 \\ \hline
Number of subjects that pass contour quality-control &  51549\\ \hline
Number of subjects for which ED surface meshes are successfully built &  51358\\ \hline
Number of subjects with viable ED surface meshes after removal of outlier meshes & 48993\\ \hline
Number of surface meshes used to create the average meshes after binning & 46917\\ \hline
Number of initial average surface meshes & 1428\\ \hline
Number of average meshes for which fibers and UVCs were successfully created & 1423\\ \hline

\end{tabular}
\label{S2_Table}
\begin{flushleft} A number of subjects or meshes were removed at certain steps of the pipeline due to reasons such as missing views, failing automatic quality-control checks, failing to produce an output etc. This table details the reasons for removal and the number of subjects or meshes remaining after these removals.
\end{flushleft}
\end{table}

\begin{figure}[!h]
\centering
\includegraphics[scale=0.59]{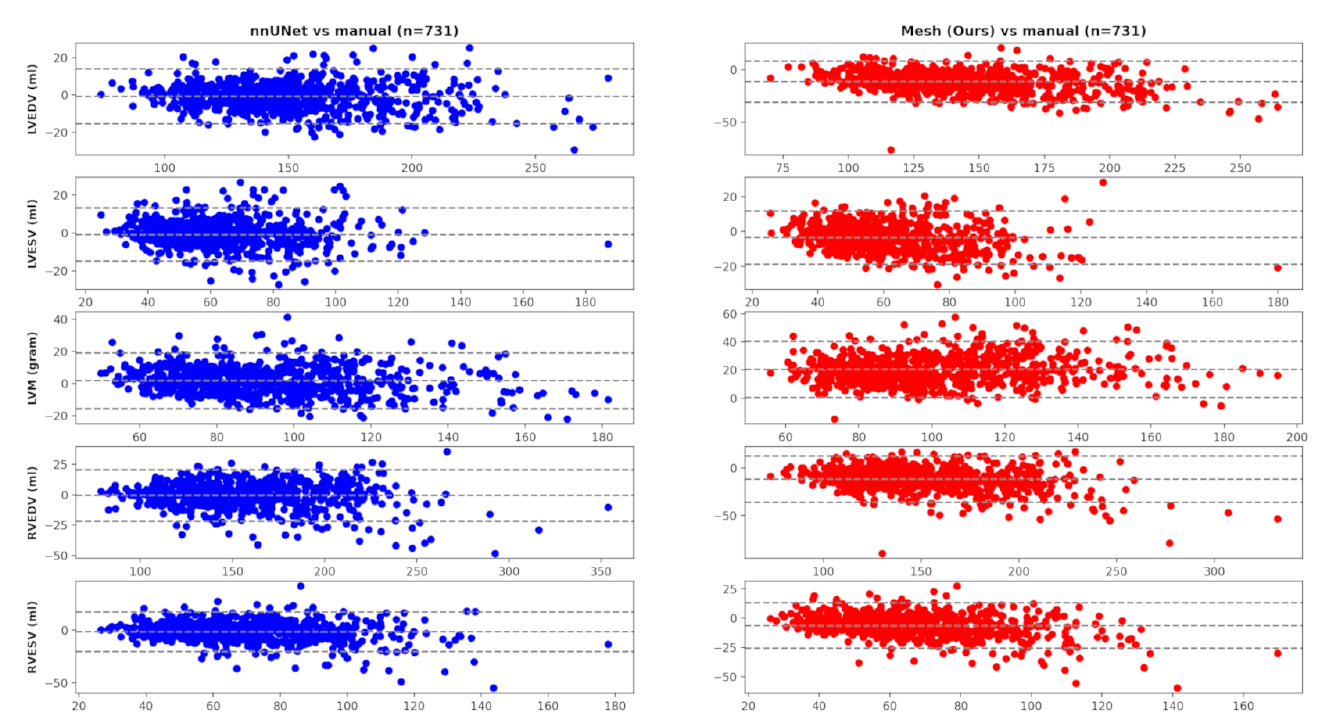}
\caption{{\bf Bland-Altman plots for phenotypes derived from the SAX view.}}
\label{S1_Fig}
\end{figure}

\section*{Funding Statement}
This work was supported primarily by the Wellcome EPSRC Centre for Medical Engineering (WT203148/Z/16/Z). Besides, SN is supported by Engineering and Physical Sciences Research Council (EP/X012603/1, EP/P01268X/1, EP/X03870X/1), British Heart Foundation (RG/20/4/34803), Foundation for the National Institutes of Health (R01-HL152256) and European Research Council (PREDICT-HF 453 (864055)). MB is supported by British Heart Foundation (PG/22/10871, PG/22/11159, PG/18/74/34077). AY is supported by Foundation for the National Institutes of Health (R01-HL121754) and Engineering and Physical Sciences Research Council (EP/Z533762/1). MS is supported by British Heart Foundation(RE/18/4/34215) and NIHR Imperial Biomedical Research Centre. PL is supported by King's British Heart Foundation (BHF) Centre of Research Excellence (RE/24/130035) and Wellcome Trust (209450/Z/17/Z).
The funders did not play any role in the study design, data collection and analysis, decision to publish, or preparation of the manuscript. There was no additional external funding received for this study.

\end{document}